\documentclass[preprint]{aastex}

\usepackage{graphicx}
\usepackage{subfigure}
\usepackage{latexsym}

\newcommand{\msun}{$M_\odot$}
\newcommand{\etal}{{et~al.}}
\newcommand{\cmsq}{cm$^{-2}$}
\newcommand{\mhi}{$M_{HI}$}
\newcommand{\nhi}{$N_{HI}$}
\newcommand{\zsun}{$Z_\odot$}
\def\dex#1{10$^{#1}$}
\def\tdex#1{$\times$10$^{#1}$}
\def\cmm#1{~cm$^{-#1}$}
\newcommand{\kms}{~km\,s$^{-1}$}
\newcommand{\vlsr}{$v_{\rm LSR}$}
\newcommand{\vlsra}{$\vert$$v_{\rm LSR}$$\vert$}

\newcommand{\HI}{\protect\ion{H}{1}}
\newcommand{\OI}{\protect\ion{O}{1}}
\newcommand{\OVI}{\protect\ion{O}{6}}
\newcommand{\CII}{\protect\ion{C}{2}}
\newcommand{\CIII}{\protect\ion{C}{3}}
\newcommand{\CIV}{\protect\ion{C}{4}}
\newcommand{\NV}{\protect\ion{N}{5}}
\def\l{$\lambda$}

\begin{document}
\title{Searching for the Intra-Group Medium in Loose Groups of Galaxies.}
\author{D.J. Pisano\altaffilmark{1}$^,$\altaffilmark{2}}
\affil{Australia Telescope National Facility, P.O. Box 76, Epping NSW 1710, 
Australia}

\author{Bart P. Wakker, Eric M. Wilcots, \& Dirk Fabian}
\affil{Astronomy Department, University of Wisconsin - Madison}
\affil{475 N. Charter St., Madison, WI 53706}

\altaffiltext{1}{NSF MPS Distinguished International Postdoctoral Research 
Fellow}
\altaffiltext{2}{email:  DJ.Pisano@csiro.au}

\begin{abstract}

We have conducted a study with the VLA, the DRAO ST, and the FUSE satellite to 
search for the intra-group medium in two loose groups of galaxies: GH 144 and GH 
158.  The VLA observations provide a census of the dense \HI\ content of these 
groups in the form of individual galaxies and free-floating \HI\ clouds as traced 
by the 21-cm \HI\ line, while the FUSE observations trace the diffuse neutral and 
hot ionized gas that may fill the intra-group medium, populate the halos of 
individual galaxies, or reside in a skin around denser, neutral clouds.  While 
nothing was detected in GH 158, in GH 144 we detected two previously unknown 
\HI-rich low-surface brightness group galaxies.  In addition, Ly-$\alpha$, 
Ly-$\beta$, \CIII\, and \NV\ were detected towards GH 144.  Using this suite of 
data, we were able to place limits on the mass of various portions of this 
group.  If virialized, GH 144 has a mass of 2$\times$10$^{12}$\msun.  Of 
that mass, 8\% lies in the individual catalogued galaxies, and no more 
than that same fraction again could lie in the dense, neutral medium as 
constrained by our VLA observations.  The absorption lines imply a 
diffuse gas with a volume density greater than 10$^{-5.2}\,$cm$^{-3}\,$ from a 
layer less than 22 kpc thick, assuming a metallicity of 0.4 \zsun.  While the 
extent of this gas is uncertain, it seems unlikely that this diffuse gas 
contributes a significant fraction of the group mass.  Given the depth of the 
absorbing material, and its separation from the nearest galaxies, it seems most 
likely that it originates from a small clump in the intra-group medium; perhaps an 
ionized high-velocity cloud, but it may be associated with one of our new \HI\
detections.  This was an ambitious first attempt to search for the IGM in emission 
and absorption, and while it was only partially successful we show what is possible 
and what more is needed for its success.

\end{abstract}

\keywords{Galaxies:  formation --- intergalactic medium --- quasars:  
absorption lines}

\section{Introduction}

The vast majority of galaxies, including the Milky Way, reside in poor
groups: collections of a few large ($\sim L_{*}$) and tens of smaller
galaxies (Tully 1987).  Long known to be the basic building blocks of
large scale structure, the importance of understanding galaxy groups has
grown with possibility that groups may contain most of the baryons in the
Universe in the hot phase of their intragroup medium (e.g. Tripp \& Savage 
2000, Tripp, Savage, \& Jenkins 2000).  Loose groups also represent a 
laboratory for the study of structure formation.  While some groups may still be 
in the process of forming themselves (Zabludoff \& Mulchaey 1998), they may also 
host ongoing galaxy formation.  Analogs to the high velocity clouds seen around 
the Milky Way may exist in other groups and represent material falling onto 
galaxies for the 
first time (see Wakker \& van Woerden 1997 and references therein for further 
discussion).  Furthermore, Blitz \etal\ (1999) and Braun \& Burton (1999, 2000) 
have argued that an important subset of high velocity clouds lie at large 
distances and, therefore, represent the large population of ``mini-halos'' 
predicted in Cold Dark Matter (CDM) models of structure formation.

The basic problem is that the baryon density in the local Universe
derived from stars, neutral hydrogen, molecular hydrogen, and hot
X-ray emitting gas represents less than a third of the baryon density
observed at high redshift (z $>>$ 2) in the Ly$\alpha$ forest (e.g.
Fukugita, Hogan, \& Peebles 1998).  A large fraction of the
``missing'' baryons could lie in a diffuse, warm intergalactic medium
with temperatures of $10^{5} - 10^{7}\,$K, as predicted by a number of
hydrodynamic simulations of galaxy and structure formation (e.g. Cen
et al. 1995, Dav\'e et al. 2001).  While such gas would be
extraordinarily difficult to detect in emission, it is evident in
absorption.  Tripp \etal\ (2000) and Tripp \& Savage
(2000) used STIS (Space Telescope Imaging Spectrograph) observations
of OVI $\lambda\lambda1032, 1038$ absorbers to suggest that most of
the baryons in the local Universe do reside in this warm component in
the intragroup medium.  Follow-up simulations by Cen et al.  (2001)
suggest that this warm intergalactic medium should be pervasive and is
a natural outcome of the growth of large scale structure.

One question that arises is whether or not there is any neutral gas
associated with the recently detected O VI absorbers in the intergalactic
medium in nearby groups; for example, does the O VI absorption arise
in the extended halos of individual galaxies, or is it associated with
the outer edges of intergalactic HI clouds?  There is a
long history of efforts to correlate Ly$\alpha$ absorption line systems with
known galaxies or intergalactic 21 cm line H~I emission with mixed
success (e.g. van Gorkom et al. 1996).  While some teams
(e.g. Lanzetta et al. 1995) claim that the majority of nearby
Ly$\alpha$ absorbers probe the extended gaseous halos of individual
galaxies, others (e.g. Tripp, Lu, \& Savage 1998, Stocke et al. 1995)
find no correlation between nearby absorption line systems and
individual galaxies.  Simulations of structure formation suggest that
absorption line systems and individual galaxies reside in the same
large-scale filaments (Dav\'e et al.  1999).  More recently, Penton,
Stocke, \& Shull (2002) found that nearly a quarter of nearby
Ly$\alpha$ absorbers arise in galaxy voids and that absorbers cluster
more weakly than galaxies. The detection of O VI absorption in nearby
groups affords us the opportunity for a sensitive search for intergalactic
H~I emission in the same groups.  Such emission may be related to the high-
velocity clouds seen around the Milky Way, and perhaps throughout the Local Group. 

In and near our Galaxy, \HI\ with velocities deviating from differential 
galactic rotation is very common.  Historically, gas clouds with large
non-rotational velocities (\vlsra$>$100\kms) have been called
``high-velocity clouds'' (HVCs).  It is now known  that these represent a variety 
of phenomena (see Wakker \& van Woerden 1997 for a review).  Some HVCs are
probably related to a galactic fountain (Shapiro \& Field 1976, Bregman 1980); 
some are tidal debris connected
to the Magellanic Stream (e.g. Putman \etal\ 2003) or other satellites 
(e.g. Lockman 2003); some may be infalling intergalactic gas (e.g. Complex
C; Wakker \etal\ 1999, Richter \etal\ 2001, Tripp \etal\ 2003); and some
may be associated with dark matter halos and be the remnants of the formation
of the Local Group (e.g. Blitz \etal\ 1999, Braun \& Burton 1999, de Heij \etal\
2002).  There is some variation between models in this last scenario.  Blitz
\etal\ (1999) suggest that {\it all} HVCs reside at distances of $\sim$ 1 Mpc and
have \HI\ masses of $\sim$10$^7$\msun, which is only a small fraction of their 
total mass.  In contrast, Braun \& Burton (1999) proposed that only compact HVCs 
(CHVCs) are at large distances.  Best fit models of this scenario by de Heij
\etal\ (2002), suggest the clouds have masses of 10$^{5.5}$ to 10$^7$\msun,
sizes of $\sim$2 kpc, and are distributed within 150-200 kpc of the Milky Way 
and M31.  

A strong test of the models proposed by Blitz \etal\ (1999), Braun \& Burton
(1999) and de Heij et al.\ (2002) is to ask ``would one expect to see similar 
\HI\ clouds in other galaxy groups?''.  While \HI\ clouds with masses predicted 
by Blitz \etal\ (1999) should be easily detected in other nearby groups, only the 
higher
mass clouds of Braun \& Burton (1999) and de Heij \etal\ (2002) would be visible.
More importantly, however, for cloud sizes of $\sim$2 kpc and typical peak \HI\
column densities of \dex{19}\,\cmm2, the average volume density is  $\sim$\dex{-
3}\,\cmm3.
At such low volume and column densities one expects that most of the hydrogen 
is ionized by the extra-galactic radiation field and thus the \HI\ represents 
the tip of the iceberg.  Such clouds could be more easily detected in absorption.  

In this paper we set out to address both the nature of the IGM in groups
similar to the Local Group and the nature of HVCs in these groups and, by 
analogy, in the Local Group.  Our approach is to search parts of two loose
groups of galaxies for HI emission associated with HVC analogs and to search
for evidence of hot gas in the same groups via FUSE spectra along lines of sight
through the same areas.  In this way, we should be able to detect both the 
neutral and ionized components of any IGM present in these two groups.  By 
obtaining both sets of data we should also be able to infer whether hot gas is
associated with the halos of individual galaxies, with clumps of material in the 
IGM potentially analogous to HVCs, or with the group as a whole.  
If there is HI emission associated with absorbing systems, then we can accurately
derive a metallicity for these objects and compare that with HVCs around the
Milky Way and throughout the Local Group.

The paper is laid out as follows.  In Section~\ref{sample}, we explain how the two 
groups were selected and describe their known properties.  Section~\ref{obs} 
discusses the VLA and FUSE observations of these groups.  We present the 
resulting data in Section~\ref{results}, and discuss the implications for the IGM 
and HVCs and our future plans in Section~\ref{disc}.

\section{Sample Selection}
\label{sample}

In order to simultaneously study dense neutral gas in emission and diffuse 
neutral and ionized gas in absorption, we required the fortunate alignment of a
galaxy hosting an active galactic nucleus (AGN) to serve as a probe
of a foreground loose group of galaxies.  Furthermore the AGN needs
to be bright enough in the ultraviolet to observe with FUSE.  We selected 
probes from Veron-Cetty \& Veron (1998) with bright far-ultraviolet 
fluxes, F$_{\lambda1000}\ge$2$\times$10$^{-14}~$ergs s$^{-1}~$
cm$^{-2}~$\AA$^{-1}$, which were within 2\degr\ of the center of
a galaxy group from the CfA group catalog (Geller \& Huchra 1983, 
hereafter GH83).  Upon individual
investigation of the resulting groups, we selected two loose groups
and two probes to observe:  GH 144 and GH 158 with Mrk 817 and Mrk 290
behind them.  The known properties of these groups are discussed below.

\subsection{GH 144 -- Mrk 817}

\begin{figure}
\plotone{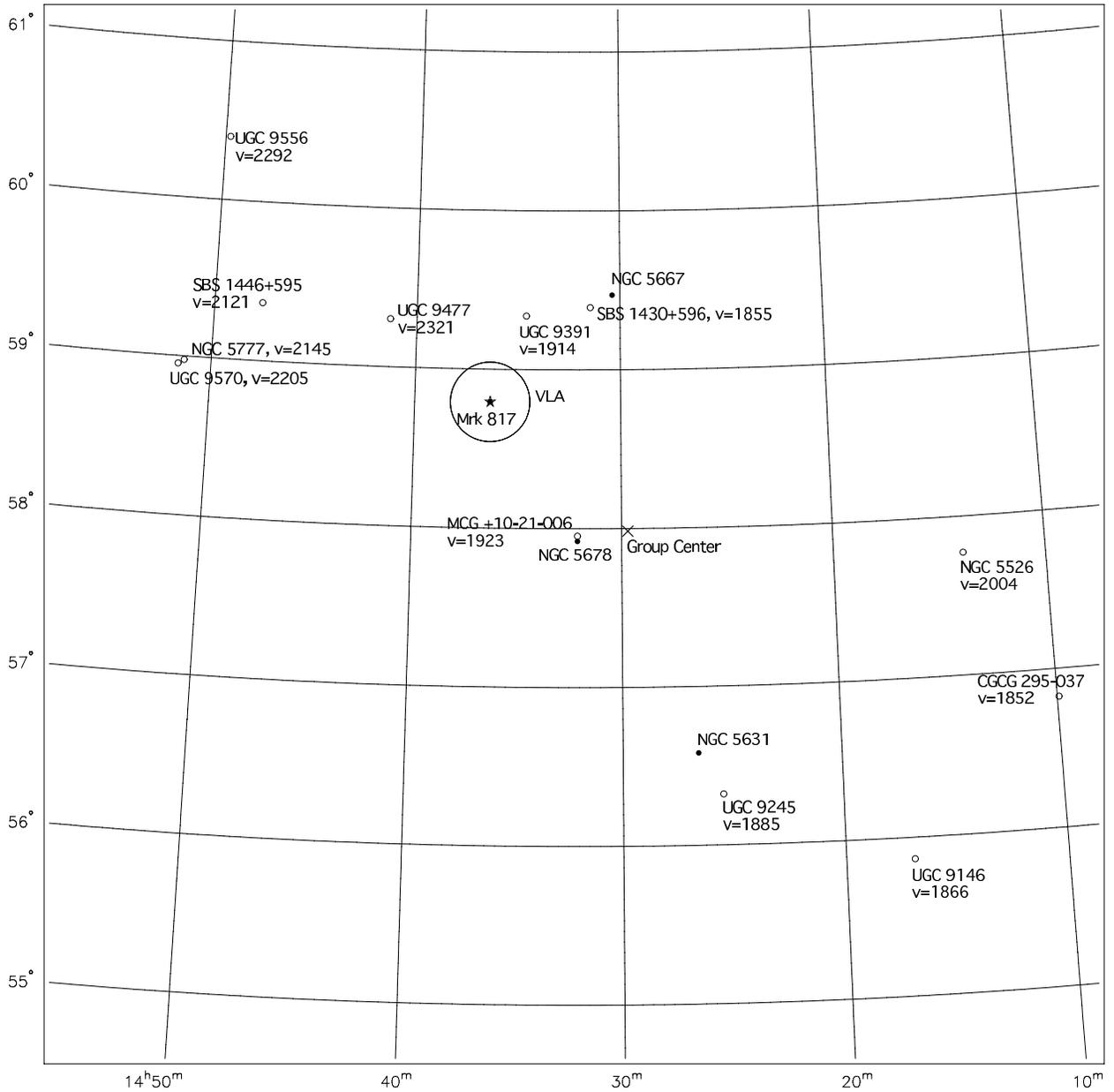}
\caption{A map of all the group galaxies (filled circles), and those galaxies 
which may be associated with GH 144 along with their heliocentric velocities (open
circles).  The star indicates the location of Mrk 817, and the large circle 
shows the area we observed with the VLA.
\label{fig:gh144map}}
\end{figure}

GH83 list GH 144 as being composed of the 3 galaxies listed in 
Table~\ref{tab:gh144grpgal}:  NGC 5631 (E), NGC 5667 (Scd), and NGC 5678 
(Sb).  GH 144 is located at a velocity of $\sim$1945 \kms; within 1000 \kms\ and 
6\degr\ of its center, an additional 11
galaxies are listed in NED\footnote{The NASA/IPAC Extragalactic Database (NED) 
is operated by the Jet Propulsion Laboratory, California Institute of Technology,
under contract with the National Aeronautics and Space Administration.} 
some of which may be part of the group.  The location of these galaxies, 
plus a few others and Mrk 817 are shown in 
Figure~\ref{fig:gh144map}.  While this is a generous range, it assures
that we have included all group members in the plot.  When corrected for 
Virgocentric infall by GH83, 
the velocity of the group is 2319 \kms.  Assuming H$_0$=65 \kms\ 
Mpc$^{-1}$, the group is at a distance of 36 Mpc.  At 36 Mpc, 1\degr\ is 
628 kpc.  Mrk 817 is located at a projected distance of $\sim$650 kpc from the
group center and 370 kpc from the nearest galaxy, UGC 9391.  This 
sightline is closer to the group center than many of the group galaxies.
Gas detected in absorption towards Mrk 817 could be associated with the
group potential or individual galaxies.  

\subsection{GH 158 -- Mrk 290}

\begin{figure}
\plotone{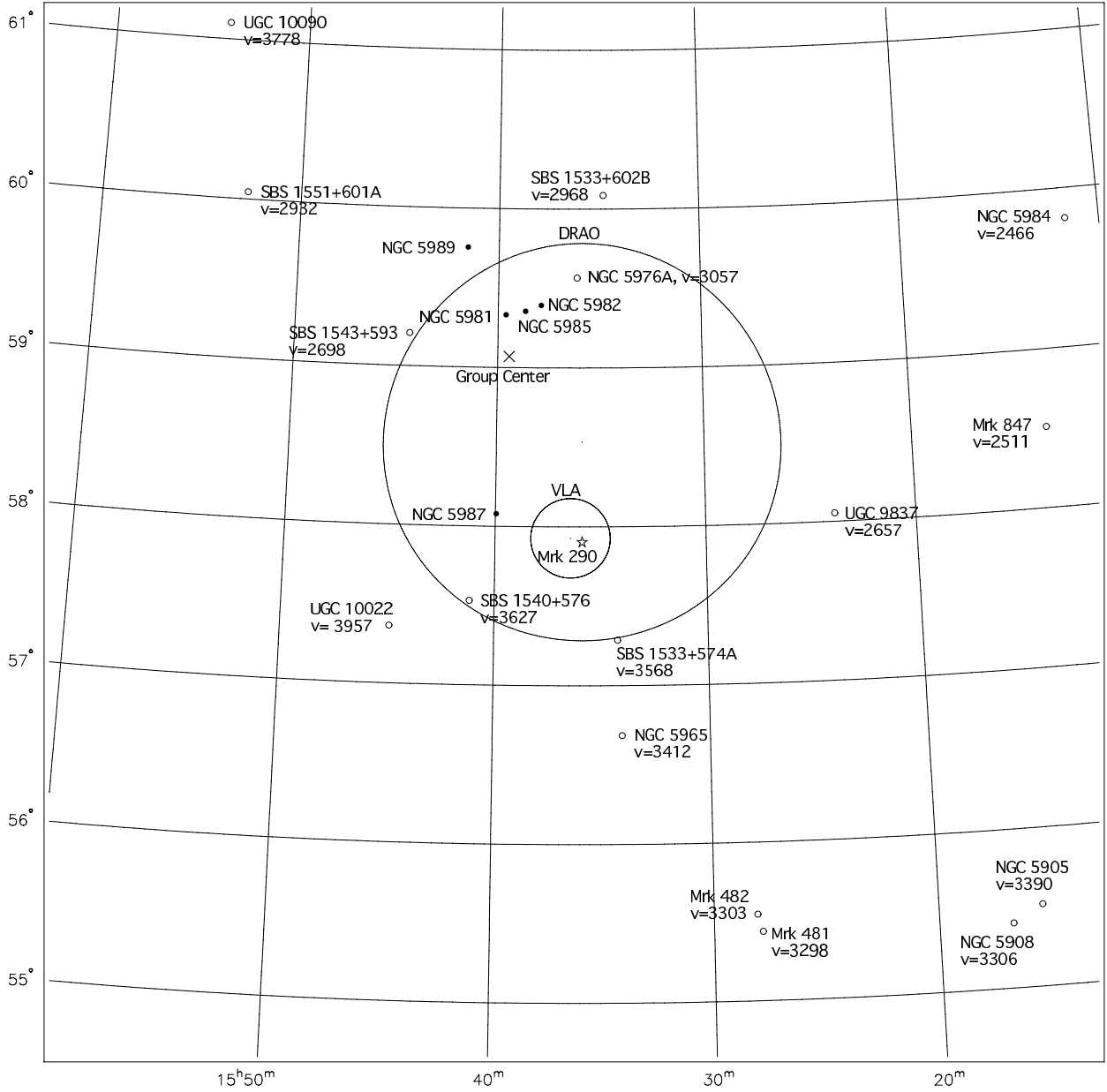}
\caption{As in Figure~\ref{fig:gh144map} but for GH 158.  
The star indicates the location of Mrk 290, the small circle shows
the area we observed with the VLA, and the large circle shows the area 
observed with the DRAO ST.
\label{fig:gh158map}}
\end{figure}

GH 158 is composed of at least the 5 galaxies listed in Table~\ref{tab:gh158grpgal} 
(GH83):  NGC 5981 (Sbc), NGC 5982 (E), NGC 5985 (Sb), NGC 5987 (Sb), and NGC 
5989 (Sc), with an additional 10 galaxies within 1000 \kms\ and 6\degr\ cataloged 
in NED.  The location of these galaxies, plus a few others and Mrk 290 are shown in 
Figure~\ref{fig:gh158map}.  With a Virgo-infall corrected velocity of 3147 
\kms, GH 158 has a distance of 48 Mpc.  At this distance, 1\degr = 837 kpc.
Mrk 290 is at a projected distance of $\sim$1 Mpc from the group
center and 480 kpc from the nearest galaxy: NGC 5987.  This sightline
probes the outskirts of the group, so it is more likely that any hot gas
seen is associated with a previously unidentified individual galaxy rather 
than with the group potential.

\section{Observations}
\label{obs}

\subsection{VLA Observations}

We observed portions of the loose groups GH 144 and GH 158 in the 21
cm line of \HI\ with the Very Large Array\footnote{The Very Large
Array (VLA) is part of the National Radio Astronomy Observatory, which
is a facility of the National Science Foundation, operated under
cooperative agreement by Associated Universities, Inc..} in its
compact D configuration.  
For GH 144, the VLA pointing is centered on Mrk 817, while for GH 158,
the pointing is between Mrk 290 and NGC 5987.  The groups were
observed on 19 and 20 August, 2000 for approximately 8.5 hours each.
The fields were observed using split intermediate-frequencies (IF) in
4 IF mode.  This yielded a velocity resolution of $\sim$5.2 \kms\ over
a total bandwidth of $\sim$600 \kms.  Flux calibration was done via
observations of 3C147 and 3C295.   Throughout the observations of the
groups, we interspersed observations of  secondary phase calibrators
about once per hour.

Calibration and reduction was carried out using AIPS in the usual
manner.  As our observations were taken almost entirely during
daylight, there was a lot of solar interference present
throughout our data.  Thus, we had to aggressively flag our data; as
much as 15\% of the data was discarded, almost all of which was on the
shortest baselines.  Each individual IF was
continuum subtracted in the uv-plane with UVLIN using  line-free
channels near the end of each cube.  The continuum-subtracted data was
then CLEANed and mapped using a robust weighting scheme (albeit close
to  natural weighting) in the AIPS task IMAGR.  The resulting data
cubes for  each IF were then glued together to form a final 600 \kms\
wide cube.  For each field we mapped a region of 1\degr\ in diameter,
out to 1\% of the  peak sensitivity of the primary beam of the VLA.
The resulting properties of the data cubes are listed in
Table~\ref{tab:hiobs}.

Our final cubes had synthesized beamsizes of 58$\arcsec$ and 1$\sigma$ 
sensitivities of 0.7 mJy/Beam and 0.5 mJy/Beam at the centers of the primary 
beams.  This translates into a
1 $\sigma$ \mhi\ sensitivity of 10$^6\,$\msun\ per channel and column density
sensitivities of  1.2$\times$10$^{18}\,$\cmsq\ per channel and
8.5$\times$10$^{17}\,$\cmsq\ per  channel for GH 144 and GH 158,
respectively.  The sensitivity of our observations is worse by a factor of two at 
a radius of 15$\arcmin$, by a factor of 4 at a radius of 21.8$\arcmin$, and is 
42$\times$ worse at 32$\arcmin$, the edge of our cubes.  For both groups we 
searched for \HI\ features not only in these cubes, but also in cubes smoothed in 
both velocity and spatially.  The smoothed cubes had noise levels reduced by up 
to 40\% of their original value.

\subsection{DRAO Observations}

Additional \HI\ observations of GH 158 were made with the Dominion
Radio Astrophysical Observatory (DRAO) Synthesis Telescope (ST) at
Penticton (Canada) in July and August 2001.  This instrument consists
of  seven 9-m radio antennae on an east-west line with a maximum
spacing  of 600-m.  For a more detailed  description of the telescope
see  Landecker \etal\ (2000).  Observations were carried out by the
staff at DRAO.  A  ``complete survey'' was made by combining
observations from 12 different configurations, with each
configuration making a full 12 hour observation of the field.  The
resulting visibility data were edited and calibrated by the observatory
staff, who also produced the initial raw data cube.

To set the flux scale of the data, we compare the fluxes of continuum
sources in the data cube with those in a ``matched continuum'' map
produced by the observatory staff.  This  ``matched continuum'' is from
a continuum channel away from the line emission and has better
determined fluxes than the line data.  Examining continuum sources in
the line-free regions of the data cube show that the continuum is flat
across the bandwidth to within 5\%.  The continuum subtracted data
cube was produced by scaling the data cube to  match the continuum
source fluxes and then subtracting the ``matched continuum'' map from
the data cube.  This is the standard procedure for continuum
subtraction of DRAO ST data.   Because of the dense sampling of the
{\it{uv}}-plane by the ST, the sidelobe level is quite low
($\le$4\%).  This fact combined with the relatively low dynamic range
of the data means that CLEANing is not necessary.

The resulting data have a beamsize of 70\arcsec\ x 59\arcsec, over a 
field of view of 2.5\degr.  The characteristics of the DRAO observations
are summarized in Table~\ref{tab:hiobs}.  The DRAO sensitivity
was 25$\times$ worse than the VLA observations of the same field.
As for the VLA data, we searched for \HI\ emission in this data cube and in 
ones which had been smoothed in velocity and spatially.  Again, the smoothed
cubes had noise levels reduced by up to 40\% of that in the original cubes.  

\subsection{FUSE Observations}

A full description of the Far Ultraviolet Spectroscope Explorer (FUSE) 
satellite and its detectors is given by Moos et
al.\ (2000) and Sahnow et al.\ (2000). Our data were taken from the survey of
Wakker et al.\ (2003). These authors describe the steps in the data reduction,
which we now summarize. First the data are screened to remove bad intervals, then
they are corrected for satellite and detector motion, extracted, and wavelength
and flux calibrated. The result is a spectrum for each of 8 detector segments.
Four of these (LiF1A, LiF2B, SiC1A and SiC2B) cover the wavelength
range near Ly$\beta$ (1025.722 \AA), but the SiC channels are much noisier than
the LiF channels. For spectra with high S/N ratio it is preferable to use only
the LiF1A data, whereas for noisier spectra the S/N ratio can be substantially
improved by combining LiF1A and LiF2B.  Wakker et al.\ (2003) further describe
how to align the FUSE spectra using \HI\ data, and how to remove the
contaminating H$_2$ lines at 1031.191 and 1032.356 \AA.  This data has been
recalibrated with the newest version (v.2.1.7) of the FUSE data pipeline, 
yielding an improved wavelength scale compared to Wakekr \etal\ (2003).  

 Mrk\,817 is relatively bright near 1030 \AA\ (a flux of 9.4\tdex{-14}
erg\,\cmm2\,s$^{-1}$\,\AA$^{-1}$ (the 16th highest in the FUSE sample) and 
two long observations are available, 76.3 and 86 ksec in length. The final 
spectrum presented by Wakker et al.\ (2003) has a signal-to-noise ratio in the 
LiF1A channel of 28.8 per 20 \kms\ resolution element near 1030 \AA\, which is the
third highest in the sample.  On the other hand, Mrk\,290 is relatively faint
(94th in the FUSE sample) and only a short (12.8 ksec) integration is 
available, yielding a final S/N ratio in the combined LiF1A+LiF2B data of just 
4.4 per resolution element.

 For Mrk\,817 HST-GHRS (the Goddard High Resolution Spectrograph on the
Hubble Space Telescope) data are also available, from HST program 6593, PI
Stocke. These data were analyzed in detail by Penton et al.\ (2000).

\section{Results}
\label{results}

\subsection{GH 158-Mrk 290}

For our observations of GH 158, we chose the VLA field so that Mrk 290 was 
near the center of the field, yet we would still have the nearest galaxy, 
NGC 5987, within the 1\degr\ field.  Unfortunately, this still places
NGC 5987 near the edge of our field, where the sensitivity has dropped off to
1\% of the peak sensitivity, and it was undetected.  To enhance our sensitivity
we Hanning smoothed our data in steps up to 42 \kms\ resolution and with a 
Gaussian filter up to 2\arcmin\ and 3\arcmin\ resolution to best match the
optical size of galaxies in the field.  While the data was up to 2.5$\times$ 
more sensitive, no \HI\ emission was seen in this or our original VLA data.  

For the DRAO pointing we chose to center the larger field between the group 
center and Mrk 290, so that 7 group galaxies lie within 
the half power radius of the primary beam.  No signature of \HI\ emission was 
present for any of the galaxies in this data.  Hanning smoothing by up to 8$\times$
in velocity and Gaussian smoothing to a resolution of 2\arcmin\ did not reveal
any emission despite improving our sensitivity by 2.5$\times$.  While two group 
galaxies, NGC 5985 and NGC 5987, have previously been detected in \HI, their 
expected fluxes of 35.1 Jy~\kms and 21.9 Jy~\kms over velocity widths of
500 \kms\ and 550 \kms\ respectively (Huchtmeier \& Richter 1989), imply an
average flux below the 2$\sigma$ level.  While the ``horns'' of
their double-horned profile may be higher than this, most of the flux would 
be at even lower significance.  The large velocity width of 
the two galaxies may also mean that not all of the emission is contained in our
bandwidth.  Clearly, more time spent observing this field, preferably with a larger 
bandwidth but similar velocity resolution, would be beneficial in detecting 
any \HI\ present.  

 Figure~\ref{fig:m290uv} shows the Ly$\beta$, Ly$\gamma$,
\OVI\l1031.926,  \OVI\l1037.617 and \CIII\l977.200 lines in the 
spectrum of Mrk 290, for a $\pm$500 \kms\ velocity range  centered around the 
average heliocentric velocity of GH158, 2850 \kms.  There is no Ly$\beta$ 
associated with GH158 with v $<$ 2800 \kms, and any absorption at higher 
velocities would be confused with Galactic \CII\ and \CII$^*$ absorption.  The 
Ly$\gamma$ and \CIII\ data are very noisy
(S/N$\sim$1), although there may be a hint of \CIII\ absorption near 2600 \kms.  
Intra-group \OVI\ \l1031.926 would fall at the velocity of geocoronal
\OI* \l1040.943 emission. Figure~\ref{fig:m290uv}, therefore, shows only the
\OVI\ data taken during orbital night. Still, intra-group \OVI\ at 2600 \kms\
is not apparently detected, with a detection
limit of 100 m\AA, or N$_{OVI}$ $<$\dex{14} \cmm2. Compared to typical
values found in Galactic high-velocity clouds ($<$log N$_{OVI}$ $>$=14.0)
this is not a very significant limit. In summary, the data quality is
too low to derive useful limits or values on the possible
intergalactic absorption associated with GH158.  A longer FUSE observation 
of Mrk 290 has been allocated time for cycle 4, which will provide better
sensitivity for a more detailed study of this group.

\subsection{GH 144-Mrk 817}
\label{gh144res}

Our VLA observations of GH 144 were centered on Mrk 817 and no 
known group galaxies lie in the field of view, however, we did detect
2 \HI\ clouds in the field.  Channel maps of these two detections are 
shown in Figures~\ref{fig:gh144d1ch} \& \ref{fig:gh144d2ch}, and the \HI\
spectra are shown in Figures~\ref{fig:gh144d1spec} \& \ref{fig:gh144d2spec}.  
The coordinates and basic properties of our detections are listed in 
Table~\ref{tab:det}.  They have been named following IAU guidelines using the 
first letter of each author's last name and their J2000 coordinates:  
PWWF J1437+5905 and PWWF J1439+5847.

As seen in Figures~\ref{fig:gh144d1ch} \& \ref{fig:gh144d2ch}, PWWF J1437+5905 has 
a peak flux greater than 6$\sigma$, and remains above 3$\sigma$ for 6 channels (31 
\kms), while PWWF 1439+5847 has a peak flux of 5$\sigma$ and spans 10 channels (52 
\kms) with a flux above 3$\sigma$ for 8 of those channels.  After correcting for 
the response of the primary beam, we find that both detections have beam--diluted 
peak column densities of 2.6$\times$\dex{19}\,\cmsq, and \HI\ masses of 
3.4$\times$\dex{9}\,\msun and 2.5$\times$\dex{9}\,\msun.  PWWF J1437+5905 lies on 
the edge of our bandpass, so its total velocity width and \HI\ mass are certainly 
higher.  Figures~\ref{fig:gh144d1opt} and ~\ref{fig:gh144d2opt} illustrate the 
probable association of both detections with optical counterparts, with unknown 
redshifts, on the Digital Sky Survey (DSS).  This suggests that both objects are 
typical low surface brightness galaxies.  Neither galaxy is cataloged in NED, but 
both of these galaxies have been detected in a shallower but wider VLA \HI\ study 
of GH 144 around the sightline to Mrk 817 conducted by McEntaffer et al. (Hibbard, 
2002, private communication).

These galaxies are both likely members of GH 144.  While PWWF J1437+5905 has a 
projected separation from the group center of 926 kpc and 288 \kms, it is 
surrounded both spatially and in velocity space by other group members.  PWWF 
J1439+5847 is also 926 kpc from the group center, but is only 100 \kms\ from the 
group velocity.  These objects are both isolated within the group residing 580 kpc 
and 800 kpc, and 290 \kms\ and 60 \kms\ from the nearest group member 
respectively.  
Finally, both of these galaxies have a projected separation of $\ge$200 kpc from 
Mrk 817, making them the closest group galaxies to any potential absorber.  

 Figure~\ref{fig:m817uv} shows the Ly$\alpha$, Ly$\beta$, Ly$\gamma$,
\OVI\l1031.926, \OVI\l1037.617 and \CIII\l977.200 lines in the UV spectrum
of Mrk 817, for a velocity range of $\pm$500 \kms\ relative to the average
heliocentric velocity of GH144, 1945 \kms.  Table~\ref{tab:m817} lists the
derived properties for these lines.  Penton et al.\ (2000) already reported
the Ly$\alpha$ absorption in the GHRS spectrum, and listed velocities of
1933 and 2097 \kms, equivalent widths of 29$\pm$13 and 135$\pm$15
m\AA, and $b$-values of 34$\pm$13 and 40$\pm$4 \kms. The corresponding
Ly$\beta$ lines are then expected to have equivalent widths of 4$\pm$3
and 21$\pm$3 m\AA.  The weaker intergalactic line is too weak too
discern in the FUSE spectrum. On the other hand, for the
higher-velocity Ly$\alpha$ line the corresponding Ly$\beta$ is clearly
detected, at 2070 \kms, with an equivalent width of 25$\pm$7 m\AA. The
absorption at slightly shorter wavelengths is likely Galactic \OVI\ at
velocities \vlsr=60--140 \kms which blends with the Ly$\beta$
line, however it may also be \OVI\ associated with the M101 group which lies 
5\degr\  (800 kpc) away.  The two Lyman lines at 2085 \kms\ lie on a curve of 
growth that
corresponds to a $b$-value of 25$\pm$10 \kms\ and log
\nhi=13.55$\pm$0.14. For the $b$-value of 40 \kms\ given by Penton et al.\ 
(2000), log \nhi\ is not substantially different (13.49$\pm$0.05).

 Any redshifted \OVI\l1031.926 near 2100 \kms\ is obscured by Galactic
\OI\l1039.230 absorption, as well as \OI\ geocoronal emission. The strong
absorption feature near 2000 \kms\ is due to \OI\ absorption in high-velocity
cloud complex~C (see e.g. Wakker 2001). The apparent broad feature in the
\OVI\l1037.617 spectrum is an artifact -- it is only seen in one of the two
observations. The detection limit in this wavelength region is $\sim$25 m\AA,
corresponding to log N$_{OVI}$ $<$13.6.

 The features at 2025 and 2125 \kms\ in the \CIII\ spectrum are measured at
18$\pm$5 and 9$\pm$5 m\AA, and are likely to be intergalactic \CIII\ in GH144. A
gaussian fit to the lines yields $b$$\sim$8 and $\sim$13 \kms. With these $b$-values, 
the corresponding column densities are log N$_{CIII}$ =12.55$\pm$0.15 and 12.14$\pm$0.20.
Despite the narrow linewidth of the low velocity component, it is also visible in the 
noisier LiF2B channel so we are confident of its reality.  
Note that these features does not align in velocity with either the Ly$\alpha$ and Ly$\beta$ 
absorption, but the two components bracket the Ly$\alpha$ component at 2097 \kms.

 Since the sightline to Mrk\,817 passes far from any galaxies in
GH144, it is likely that photo-ionization by the extragalactic
background is the dominant source of ionization. Savage et al.\ (2002)
calculated a photo-ionization model using CLOUDY (Ferland 2003 and references
therein) 
assuming a standard extragalactic radiation field composed of QSOs and the cosmic 
microwave background for an \OVI\ absorber at slightly higher redshift, but with a 
similar value for \nhi\ (see their Fig.~7).  CLOUDY estimates the column density 
and volume density of
hydrogen (neutral plus ionized) based on fixing \nhi, N$_{OVI}$, and metallicity for a
given radiation field.  Using the Savage \etal\ model and assuming a 
metallicity of 0.4 \zsun, the \OVI\ detection limit for GH144 toward Mrk\,817 
implies an ionization parameter $\log U<-1.1$ and a total hydrogen volume density 
$\log n_H>-5.2$.  Using the derived column density and volume density, the model yields 
a depth, L, for the absorbing gas of $<$22 kpc.  The predicted \OVI\ column density, and 
inferred depth, scales non-linearly with the assumed metallicity, so that if a
metallicity of 0.1 \zsun\ is assumed, the \OVI\ detection limit implies
$\log n_H>-5.6$, and L $<$ 150 kpc.  

 \NV\l1238.821 at a velocity of 2041 and 2106 \kms\ may be detected, with
an equivalent width of 15$\pm$4 and 13$\pm$4 m\AA. Using the fitted linewidth of
$b$=60 \kms, this would imply a column density of $\log_{NV}$=13.1$\pm$0.2. 
This would imply $\log U \sim\ -0.8$. On the other hand,
the upper limit set by the \OVI\ non-detection implies log $U<-1.1$,
which yields $\log_{NV}<$12.9. Since the \NV\ detection is very
tentative this is not incompatible. Better data would obviously be
useful.

 The \CIV\ lines are unfortunately not covered by the GHRS spectrum. At
$\log U=-1.1$ the prediction is log N$_{CIV}$ $\sim$13.2. Such a line would be
easily detectable (optical depth $\sim$1) and since Mrk\,817 is bright, a high
signal-to-noise spectrum can easily be obtained with HST. An accurate
measurement of N$_{CIV}$ would allow the determination of the density of the
absorbing gas.

\section{Discussion \& Conclusions}
\label{disc}

\subsection{The mass of the IGM}

The main hope of this study was to find HI emission associated with the diffuse 
IGM that produces UV absorption lines in both groups:  none was found.  Our only 
HI detections were of two galaxies in the GH 144 group.  We can use the 
non--detection of \HI\ and the detection of the IGM in absorption, along with a few 
assumptions regarding the size of the group, to try and place limits on the total amount of 
gas in the IGM of GH 144.  In particular we will examine what fraction of the total mass 
of each group could be baryonic and residing in the IGM.  Because we lack high--quality 
FUSE data for GH 158, we will omit it from this excercise. 

The first step is to determine what the total masses of each group are based on 
the assumption that the groups are virialized.  While it is very likely that most 
loose groups have not yet fully collapsed, let alone virialized, as evidenced by 
the presence of substructure
in many groups (e.g. Zabludoff \& Mulchaey 1998), this is the only method for 
determining a total group mass based on the dynamics of the group.  Taking the 
Virial theorem from Binney \& Tremaine (1987):

\begin{equation}
\label{eq:mvirial}
M_{virial} = \frac{\sigma^2 r_g}{G}
\end{equation}

where $r_g$ is the gravitational radius of the group and $\sigma$ is the three-
dimensional velocity dispersion of the group.  As these values are not equal to 
the observable values, we must convert the equation to use values we can 
measure:  the radial velocity dispersion and the group radius.  In the former 
case, the square of the three-dimensional velocity dispersion is traditionally assumed 
to be three times the square of the radial velocity dispersion (i.e. the velocity 
dispersion is assumed to be isotropic):  $\sigma^2 \, = 3\times \, \sigma_r^2$.  
Relating an observable radius to the gravitational radius is a bit more uncertain, 
but Binney \& Tremaine (1987) state that for ``many simple stellar systems'', the 
radius which contains half the mass is 40\% of the gravitational radius:  $r_{med} 
= 0.4 r_g$.  Therefore, Equation~\ref{eq:mvirial} becomes:

\begin{equation}
\label{eq:virial}
M_{virial} = \frac{7.5 \sigma_r^2 r_{med}}{G}
\end{equation}

where $\sigma_r$ is the radial velocity dispersion, and we assume $r_{med}$ is 
half of the maximum separation between group galaxies.  Taking the velocity 
dispersion from GH83, we have the following values for $r_{med}$ and 
$\sigma_r$:  925 kpc and 36 \kms.  These values are calculated by GH83 using only 
those galaxies they listed as group members (i.e. the bright group galaxies).  
As a result of not accounting for the fainter population of galaxies likely 
present in the group, the group size and especially the velocity dispersion as 
listed by GH83 are not as accurate as they could otherwise be.  Keeping this 
caveat in mind, we find that
M$_{virial}$ is 2$\times$10$^{12} (\sigma_r/36 km s^{-1})^2$\msun\ for GH 144.  
Is this mass mostly in individual galaxies or is it spread throughout the group?

To answer this question, we use both the virial theorem and the dynamical mass 
equation:

\begin{equation}
\label{eq:mdyn}
M_{dyn} = \frac{V_{rot}^2 R_{HI}}{G}
\end{equation}

where V$_{rot}$ is the rotation velocity of the galaxy (corrected for inclination) 
and R$_{HI}$=1.7$\times$R$_{25}$, where R$_{25}$ is the radius of the galaxy 
at the 25 mag arcsec$^{-2}$ isophote.  The factor of 1.7 is an average from a 
sample of spiral galaxies studied by Broeils \& Rhee (1997).  
These values for each large galaxy in the group come from the Lyon-Meudon 
Extragalactic Database (LEDA) and are listed in Table~\ref{tab:gh144gals}.
Summing the total masses of each galaxy, 
we find that the mass in galaxies is 5$\times$10$^{11}$\msun.  This is 25\% of 
the virial mass of each group; it will be even less if the true group velocity 
dispersion is higher.  Is the remaining the group mass in the form of 
diffuse gas or is it dark matter?  

It is at this point where we use our observations to place limits on the mass of 
the IGM in GH 144.  This can be done for the dense, yet diffuse, 
neutral gas traced by the \HI\ 21-cm emission and the lower density diffuse 
ionized gas traced by the FUSE absorption line data.  We must make assumptions 
regarding the area over which the IGM is spread and its filling fraction over that 
area in combination with our column density sensitivities from the 21-cm and UV 
observations.  For the \HI\ observations, we have no a priori information to constrain 
the filling factor, so for starters we will leave it in our results explicitly.  In 
Table~\ref{tab:hiobs} the column density sensitivity of our VLA observations is listed.  
Assuming that the areas we observed are typical of the group as a whole (in that 
the \HI\ is uniformly distributed), taking a 5$\sigma$ detection limit at the 
center of the primary beam, and integrating over an area with a radius equal to 
r$_{med}$ while explicitly listing the dependence on the number 
channels our limits are:

\begin{equation}
\label{eq:high144}
M_{HI}(GH 144) \le 1.3\times10^{11}\times \sqrt{n}\times f
\end{equation}

where $n$ is the number of channels the signal is distributed across and $f$ is 
the filling fraction.  Assuming that $f = 1$ and taking the velocity dispersion
of the group (36 \kms) the mass limit becomes 3$\times$\dex{11}\msun, or less
than 67\% of the mass in individual galaxies and less than 15\% of the virial
mass of the group.  With the assumptions regarding the filling factor for the group 
and the upper limit on the \HI\ column density, this is a highly uncertain value.
Nevertheless, the presence of only weak Lyman-$\alpha$ absorption implies a very low 
filling fraction of dense neutral gas in the group.  Some of the diffuse gas in these 
groups may be hot and ionized, so we must look to our UV absorption line data for limits 
on its contribution to the total mass of the group.

To obtain the total potential gas mass of the groups, we must carry out the same 
exercise for the diffuse warm and hot ionized gas traced by the UV absorption 
lines as well.  In Section~\ref{gh144res}, we discuss how we determine the density 
and size scale for the hot gas in GH 144.  The \OVI\ line constrains the density to 
be greater than 10$^{-5.2}\,$cm$^{-3}$ for a metallicity of 0.4 \zsun.  This volume 
density, combined with the column density of hydrogen, provides a size scale of 
$<$22 kpc.  So, if this absorbing gas comes from a cloud with a 22 kpc diameter and 
a filling fraction of 0.5 (based on the number of groups with absorbing gas 
discussed in Wakker \etal, in preparation), then the mass of the cloud is 
4$\times$10$^5$\msun; a negligible fraction of the group mass.  If the metallicity 
of the gas was only 0.1 \zsun, then the size scale increases to $<$150 kpc, and the mass 
of a cloud with this size is $\sim$10$^7$\msun; still a negligible fraction of the group 
mass.  

What if the absorbing gas comes from a thin sheet of hot gas spread 
throughout the group?  Taking a depth of 22 kpc, and a diameter of 1.85 Mpc, 
then we get a mass of 4$\times$10$^9$\msun\ for our original volume density.  
In any case, in order to increase the mass contribution of this gas to the group 
mass we must increase the density, but this would imply a smaller region from 
which the absorption originates.  While this is just one line of sight through the 
group, it seems unlikely that this absorbing gas contributes any 
significant mass to the IGM.  Multiple lines of sight through this group and further
studies of other groups similar to GH 144 will help determine if this is a atypical 
line-of-sight, or if more groups lack diffuse gas.  The absence of a significant 
amount of warm to hot diffuse gas in the IGM of GH 144 is contradictory to both the 
predictions of models (e.g. Cen \etal\ 2001, Dav\'e \etal\ 2001) and observations 
towards other groups (e.g. Tripp \& Savage 2000, Tripp \etal\ 2000).  Instead, it 
appears that the majority of the mass outside of galaxies in GH 144 can not be
identified as diffuse neutral or hot gas, but may have to accounted for as
dark matter.

\subsection{What is the source of the UV absorption lines?}

As discussed in the Introduction, it is a matter of some debate whether Lyman-
$\alpha$ and \OVI\ absorbers originate in the extended gaseous halos of 
galaxies (e.g. Lanzetta \etal\ 1995, Savage \etal\ 2003), as the ionized layer 
of intergalactic
\HI\ clouds analogous to HVCs (e.g. Sembach \etal\ 2000), from the intra-group 
medium (e.g. Tripp \& Savage 2000, Tripp \etal\ 2000), or as part of large-scale 
filaments (e.g. Dav\`e \etal\ 1999).  One goal of this project was to examine the
relation between \HI\ emission and UV absorption line systems to address this question. 

Since the UV data for Mrk 290 is not of particularly high quality, we
will focus our discussion on the data towards Mrk 817 through the GH
144 group.  Ly-$\alpha$,  Ly-$\beta$, and \CIII\ lines along this
sightline as discussed above, with the more highly ionized \CIII\
bracketing one of the two Lyman-$\alpha$ lines.  The lower velocity, weaker 
Lyman-$\alpha$ line has no other associated lines, but this may be due
to our lack of sensitivity.  The system of lines at $\sim$2100 \kms\ is
consistent with what was seen by Wolfe \& Prochaska (2000a) in Damped 
Lyman-$\alpha$ protogalaxies where the more highly ionized lines lie 
outside those of lower ionization, so while the lines do not overlap 
they likely originate in the same gravitational potential.  Wolfe \& Prochaska (2000b)
interpret this  scenario arising from a line-of-sight passing through
a neutral disk within a hot, collapsing halo.  Their tests of
specific models were inconclusive, however.  We may be seeing something
similar to their scenario with hot gas condensing onto a neutral cloud in the
intra-group medium, but this is certainly not the only explanation.  

The system of absorption lines towards Mrk 817 is not centered on the velocity of
the group (1945 \kms), but is offset by $\sim$100 \kms\ to higher velocities.  
If these lines are associated with intra-group gas, this may 
indicate the presence of substructure in GH 144, so that the group may not yet 
be virialized.  There is no \HI\ emission directly associated with this 
absorption-line system.  

Another possibility comes about from examining the size scale of the absorbing 
material.  With the inferred size scales of the gas being only 22 kpc, this 
implies we are probing the outskirts of the halo of a galaxy, or part of a 
clumpy IGM.  For example, an IGM filled with highly ionized HVCs.  
In the former case, the closest galaxies to this sightline are PWWF J1437+5905 
and PWWF J1439+5847 which lie about 200 kpc away in projection.  If the UV 
absorption lines are associated with gas in the halo of either galaxy, then the 
halos must be very extended.  In velocity space, however, the absorbers have 
velocities that are consistent with the velocity width of PWWF J1439+5847.  
The Ly-$\alpha$ lines are offset by -112 \kms\ and +52 \kms\ offset while the 
\CIII\ is -20 \kms\ and +71 \kms\ offset.  Compare this to 
the velocity width at 20\% of the peak value of 60 \kms, and it would not be 
completely unreasonable to believe that they may be associated.  If they are, 
then the total mass of the halo of PWWF J1439+5847 would be only 
4$\times$\dex{10}\msun, about 10$\times~$ the \HI\ mass of the galaxy.  
Given the huge physical separation, it may be more reasonable to assume that 
the absorbing gas is instead a clump associated with the IGM or the large-scale 
filament in which the group resides (e.g. Rosenberg \etal\ 2003).  This clumpy IGM 
could represent low mass, mostly ionized, HVCs spread throughout the group.  
These would be of lower mass than would be inferred based on the models of 
Blitz \etal\ (1999) and Braun \& Burton (1999), but may otherwise be analogous.

More sensitive \HI\ observations and the ability to probe the absorbing gas in 
multiple sightlines through a single group should provide better constraints on 
the competing models for the source of the UV absorption lines.

\subsection{Future Possibilities}
\label{conc}

Overall our study has been an ambitious first attempt to study the nature of the
intra-group medium in both emission and absorption, which has been only partially
successful.  It has yielded some interesting insights into the relation of the 
diffuse warm, absorbing gas with the denser, cooler, neutral, emitting gas in a 
loose group.  While it is difficult to learn much from the study of just a couple 
groups, Wakker \etal\ (2004, in preparation) have identified other galaxy groups with background AGN suitable for this type of study.  The most fundamental 
limitations to our study come from the poor \HI\ sensitivity to diffuse, low 
column density 
gas and the small area mapped with a single VLA pointing, and the lack of multiple 
sightlines to background AGNs through these groups.  In the former case large area 
maps made with single dish radio telescopes equipped with multibeam receivers 
(such as the Parkes Multibeam and the Arecibo ALFA receiver currently under 
construction) should provide better constraints.  In the latter case, we must wait 
for the next generation of space-based instruments, starting with the Cosmic Origins 
Spectrograph (COS) on HST.  Such instruments will start to revolutionize this field.  
They will permit fainter AGNs to be used as lightbulbs for absorption line studies 
allowing for more groups to be studied and even for studies of multiple sightlines 
through individual groups.  As an example, COS will be an order of magnitude more 
sensitive than FUSE.  Future instrumentation on UV/Optical space telescopes should 
be even better.  When such instruments become available, we hope that our work will 
help serve as a template for future multiwavelength studies of groups of galaxies.

\acknowledgements  The authors would like to thank the staff at the VLA for
their assistance in the reduction of the data.  We thank John Hibbard
for providing us with information on the McEntaffer \etal\ results before 
publication.  We also thank Hugo van Woerden for a helpful referee's report.
Some of the data presented in this paper was obtained from the 
Multimission Archive at the Space Telescope Science Institute (MAST); STScI is
operated by the Association of Universities for Research in Astronomy,
Inc., under NASA contract NAS5-26555. Support for MAST for non-HST
data is provided by the NASA Office of Space Science via grant
NAG5-7584.  This research  has made use of the NASA/IPAC Extragalactic
Database (NED) which is operated  by the Jet Propulsion Laboratory,
California Institute of Technology, under  contract with the National
Aeronautics and Space Administration.  The Digitized Sky Surveys were
produced at the Space Telescope Science  Institute under
U.S. Government grant NAG W-2166. The images of these  surveys are
based on photographic data obtained using the Oschin Schmidt
Telescope on Palomar Mountain and the UK Schmidt Telescope. The plates
were processed into the present compressed digital form with the
permission  of these institutions.  The Second Palomar Observatory Sky
Survey (POSS-II) was made by the California Institute of Technology
with funds from the National Science Foundation, the National
Geographic Society, the Sloan Foundation, the Samuel Oschin
Foundation, and the Eastman Kodak Corporation.  We have made use of the
LEDA database (http://leda.univ-lyon1.fr).  D.J.P. and D.F. acknowledge 
partial support for this project from Wisconsin Space Grant Graduate Fellowships.  
D.J.P. acknowledges generous support from an NSF MPS Distinguished International
Postdoctoral Research Fellowship, NSF grant AST0104439.   B.P.W. was
supported by NASA grants NAG5-9024, NAG5-9179, and NAG5-8967.  E.M.W.
was kindly supported by NSF grant AST 98-75008.

\clearpage

\begin{deluxetable}{lcccc}
\tablecolumns{5}
\tablewidth{0pc}
\tablecaption{GH 144 Group Galaxies\label{tab:gh144grpgal}}
\tablehead{\colhead{Galaxy} & \colhead{$\alpha$ (J2000)} & \colhead{$\delta$ (J2000)} & 
\colhead{V$_\odot$} & \colhead{Hubble Type}  \\
\colhead{}&\colhead{h:m:s}&\colhead{d:m:s}&\colhead{\kms}&\colhead{}}
\startdata
NGC 5631 & 14:26:33.3 & 56:35:00 & 1979 & E    \\
NGC 5667 & 14:30:24.9 & 59:28:16 & 1943 & Scd  \\
NGC 5678 & 14:32:05.4 & 57:55:13 & 1922 & Sb   \\
\enddata
\tablecomments{All data is taken from NED.}
\end{deluxetable}

\begin{deluxetable}{lcccc}
\tablecolumns{5}
\tablewidth{0pc}
\tablecaption{GH 158 Group Galaxies\label{tab:gh158grpgal}}
\tablehead{\colhead{Galaxy} & \colhead{$\alpha$ (J2000)} & 
\colhead{$\delta$ (J2000)} & \colhead{V$_\odot$} & \colhead{Hubble Type}  \\
\colhead{}&\colhead{h:m:s}&\colhead{d:m:s}&\colhead{\kms}&\colhead{}}
\startdata
NGC 5981 & 15:37:53.3 & 59:23:34 & 1764 & Sbc \\
NGC 5982 & 15:38:39.7 & 59:21:20 & 2904 & E   \\
NGC 5985 & 15:39:37.7 & 59:19:57 & 2517 & Sb  \\
NGC 5987 & 15:39:58.1 & 58:04:56 & 3010 & Sb  \\
NGC 5989 & 15:41:32.5 & 59:45:19 & 2878 & Sc  \\
\enddata
\tablecomments{All data is taken from NED.}
\end{deluxetable}

\clearpage

\begin{deluxetable}{lccc}
\tablecolumns{4}
\tablewidth{0pc}
\tablecaption{HI Observation Properties\label{tab:hiobs}}
\tablehead{\colhead{} & \colhead{GH 144 - Mrk 817} & \colhead{GH 158 - Mrk 
290} & \colhead{GH 158 - Mrk 290}}
\startdata
Telescope 	  & VLA & VLA 	& DRAO ST \\
Configuration 	  & D 	& D 	& \nodata \\
Observation Dates & 2000 Aug. 19,20 & 2000 Aug. 19,20 & 2001 July \& Aug. \\
Integration Time (hr) & 8.1 & 8.1 & $\sim$144 \\
Beam Size (arcsec)  & 58 & 58 & 70$\times$59 \\
Half Power Diameter FOV (deg.) & 0.5 & 0.5 & 2.5 \\
Channel Width (\kms) & 5.2 & 5.2 & 3.3 \\
Velocity Coverage (\kms) & 1642 - 2254 & 2593 - 3207 & 2474 - 3323 \\
Sensitivity\tablenotemark{a}~(mJy/Bm) & 0.7 & 0.5 & 9.8 \\
Sensitivity\tablenotemark{a}~(\dex6 \msun) & 1.0 & 1.4 & 22 \\
Sensitivity\tablenotemark{a}~(\dex{18}\cmsq) & 1.2 & 0.85 & 9 \\
\enddata
\tablenotetext{a}{The 1$\sigma$, 1 channel RMS noise at the center of each data 
cube.}
\end{deluxetable}

\begin{deluxetable}{lcc}
\tablecolumns{3}
\tablewidth{0pc}
\tablecaption{Properties of GH 144 \HI\ Detections\label{tab:det}}
\tablehead{\colhead{} & \colhead{PWWF J1437+5905} & \colhead{PWWF 
J1439+5847}}
\startdata
$\alpha$ (J2000)                 & 14:37:02   & 14:39:04   \\
$\delta$ (J2000)                 & +59:05:55  & +58:47:29  \\
V$_\odot$ (\kms)  		 & 2233       & 2045       \\
W$_{20}$ (\kms)                & 34\tablenotemark{a} &   60   \\
\mhi\ (\dex{9}\msun)             & 3.4\tablenotemark{a} & 2.5        \\
Peak \nhi\ (\dex{19}\cmsq)        & 2.6        & 2.6        \\
Separation (kpc)\tablenotemark{b}& 200        & 221        \\
Separation (\kms)                & 288\tablenotemark{a} & 100        \\
\enddata
\tablenotetext{a}{These are lower limits since this object extends beyond our 
observed bandpass.}
\tablenotetext{b}{The projected radial separation between Mrk 817 and each 
\HI\
detection, assuming they lie at the group distance of 36 Mpc (1\arcmin\ = 10.5
kpc).}   
\end{deluxetable}

\clearpage

\begin{deluxetable}{lccccc}
\tablecolumns{6}
\tablewidth{0pc}
\tablecaption{Mrk 817 UV data\label{tab:m817}}
\tablehead{\colhead{Ion}&\colhead{V$_\odot$}&\colhead{W$_\lambda$}
&\colhead{b}&\colhead{log N}\\
\colhead{}&\colhead{\kms}&\colhead{m\AA}&\colhead{\kms}&\colhead{\cmsq}}
\startdata
Lyman-$\alpha$ & 1933 &  29$\pm$13 & 34$\pm$13 & 12.76$\pm$0.24 \\
Lyman-$\alpha$ & 2097 & 135$\pm$15 & 40$\pm$4  & 13.55$\pm$0.14 \\
Lyman-$\beta$  & 2070 &  25$\pm$7  & 25$\pm$10 & 13.55$\pm$0.14 \\
\NV\l1238.821  & 2041 &  15$\pm$4  & 30$\pm$8  & 12.84$\pm$0.13 \\
\NV\l1238.821  & 2106 &  13$\pm$4  & 30$\pm$8  & 12.79$\pm$0.15 \\
\CIII\l977.20  & 2025 &  18$\pm$5  &  8$\pm$3  & 12.55$\pm$0.15 \\
\CIII\l977.20  & 2116 &   9$\pm$5  & 13$\pm$5  & 12.14$\pm$0.20 \\
\OVI\l1037.617 & \nodata & $<$21\tablenotemark{a} & 30\tablenotemark{b} & $<$13.53\tablenotemark{a} \\
\tablenotetext{a}{This is a 3$\sigma$ upper limit using the assumed b-value.}
\tablenotetext{b}{This is an assumed value based on the approximate widths of the other
absorption lines.}
\enddata
\end{deluxetable}

\begin{deluxetable}{lccc}
\tablecolumns{4}
\tablewidth{0pc}
\tablecaption{GH 144 Galaxy Properties\label{tab:gh144gals}}
\tablehead{\colhead{Galaxy} & \colhead{R$_{25}$\tablenotemark{a}} 
& \colhead{V$_{rot}$\tablenotemark{b}} & \colhead{M$_{dyn}$\tablenotemark{c}}\\
\colhead{}&\colhead{kpc}&\colhead{\kms}&\colhead{\msun}}
\startdata
NGC 5631 & 10.5 & 190\tablenotemark{d} & 1.5$\times$10$^{11}$\\
NGC 5667 & 8.9 & 120 & 5.1$\times$10$^{10}$\\
NGC 5678 & 17.4 & 210 & 3.1$\times$10$^{11}$\\
\enddata
\tablenotetext{a}{from LEDA and assuming that all galaxies lie at the group 
distance.}
\tablenotetext{b}{from LEDA}
\tablenotetext{c}{calculated using Equation~\ref{eq:mdyn}.}
\tablenotetext{d}{calculated assuming V$_{rot}$=W$_{20}$/2, with W$_{20}$ 
from LEDA.}
\end{deluxetable}

\clearpage

\begin{figure}
\epsscale{0.3}
\plotone{pisano.fig3.ps}
\caption{This figure shows the Ly$\beta$, Ly$\gamma$, \OVI\l1031.926, 
\OVI\l1037.617 and \CIII\l977.200 lines, for a $\pm$500 \kms\ velocity range 
centered around the average heliocentric velocity of GH158, 2850 \kms.  Flux is 
in units of 1$\times$10$^{-14}~$erg \cmsq\ s$^{-1}$ \AA$^{-1}$.  The labels at 
the bottom identify interstellar lines, with single numbers giving the J-level of 
H$_2$ absorption lines.  The region between the ``$<$'' and ``$>$'' is unreliable 
because of a detector flaw.  The horizontal line represents the continuum which 
was fitted over a region much larger than that displayed here.  The $\oplus$ symbol
represents geocoronal lines.  
\label{fig:m290uv}}
\end{figure}

\begin{figure}
\epsscale{0.7}
\plotone{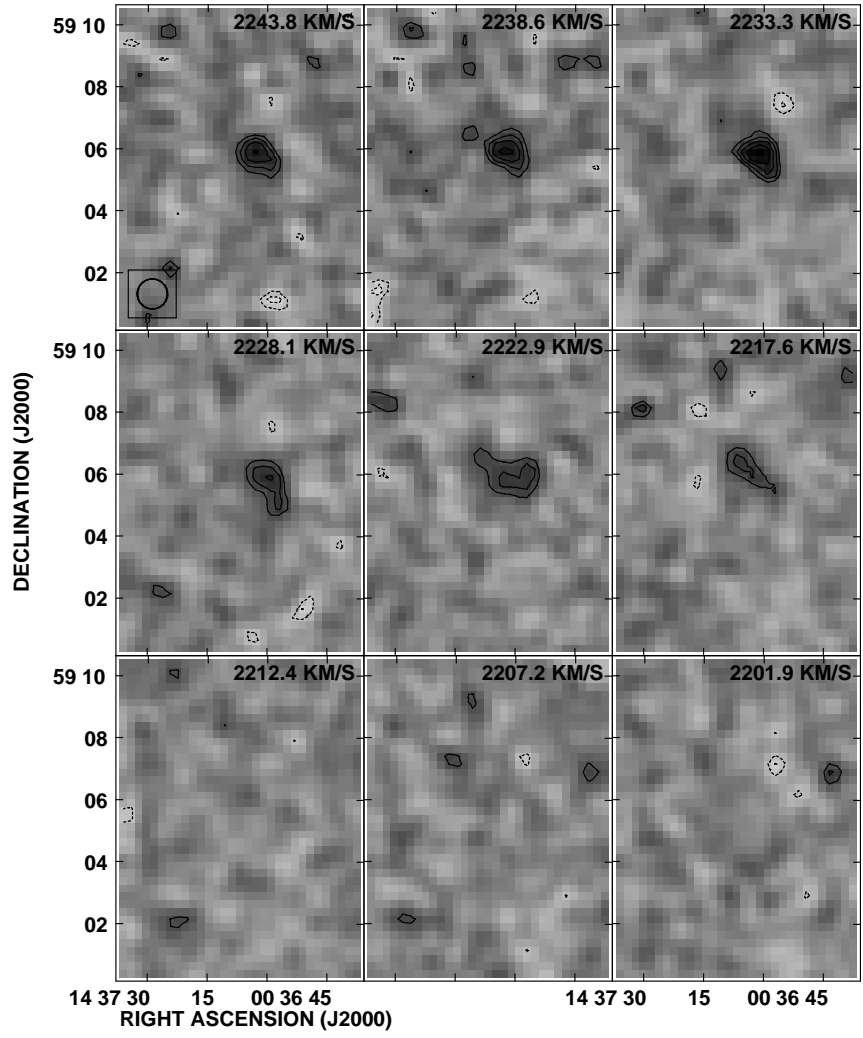}
\caption{The channel maps of PWWF J1437+5905 in the GH 144 VLA field.  
Contours are at
-4$\sigma$, -3$\sigma$, 3$\sigma$, 4$\sigma$, 5$\sigma$, 6$\sigma$ where 
$\sigma$ 
is the RMS noise (= 0.7 mJy/beam).
\label{fig:gh144d1ch}}
\end{figure}

\begin{figure}
\epsscale{0.7}
\plotone{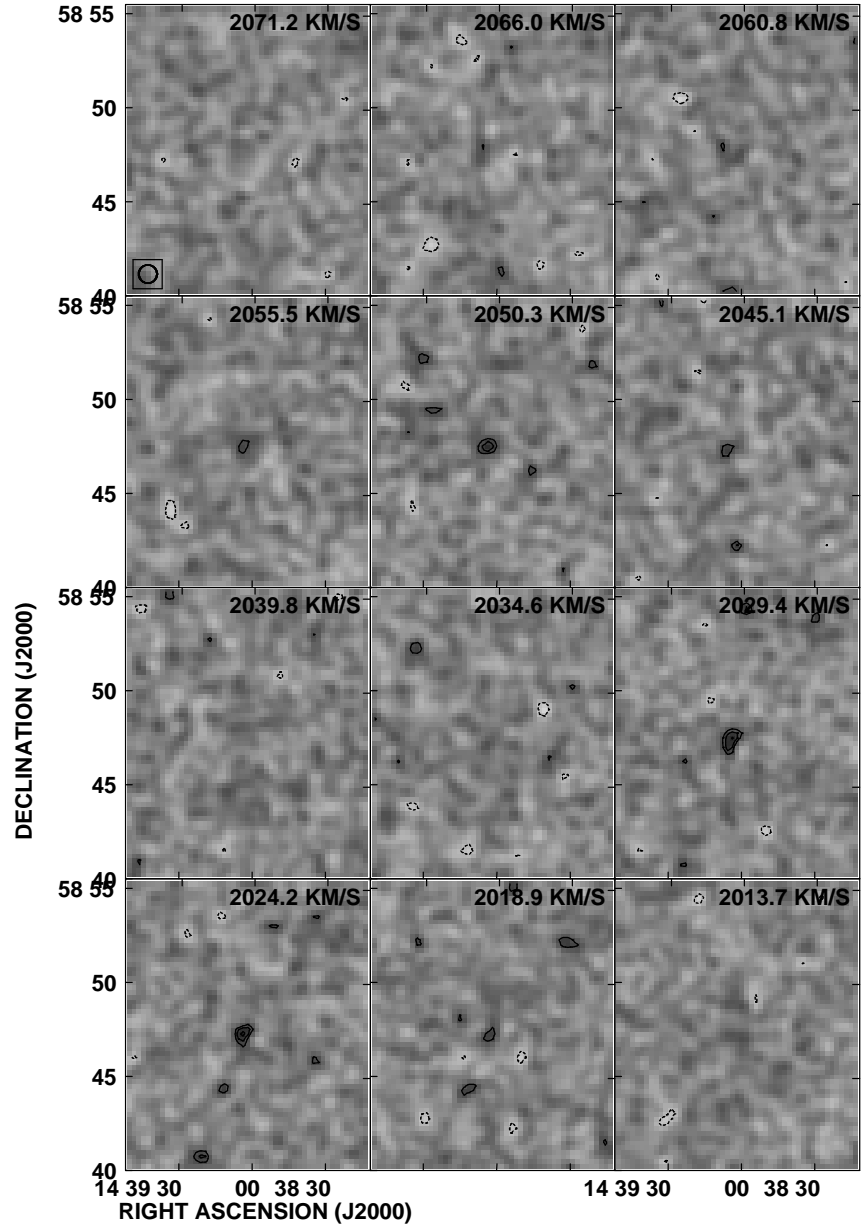}
\caption{The channel maps of PWWF J1439+5847 in the GH 144 VLA field.  
Contours are at
-3$\sigma$, 3$\sigma$, 4$\sigma$, 5$\sigma$ where $\sigma$ is the RMS noise 
(= 0.7 mJy/beam).
\label{fig:gh144d2ch}}
\end{figure}

\begin{figure}
\epsscale{0.7}
\plotone{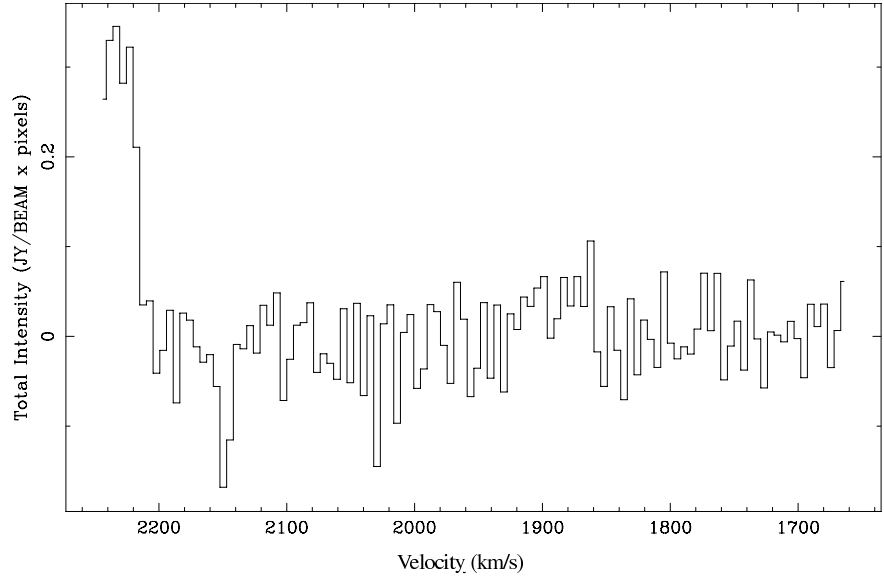}
\caption{An \HI\ spectrum of PWWF J1437+5905 after primary beam correction.  
\label{fig:gh144d1spec}}
\end{figure}

\begin{figure}
\epsscale{0.7}
\plotone{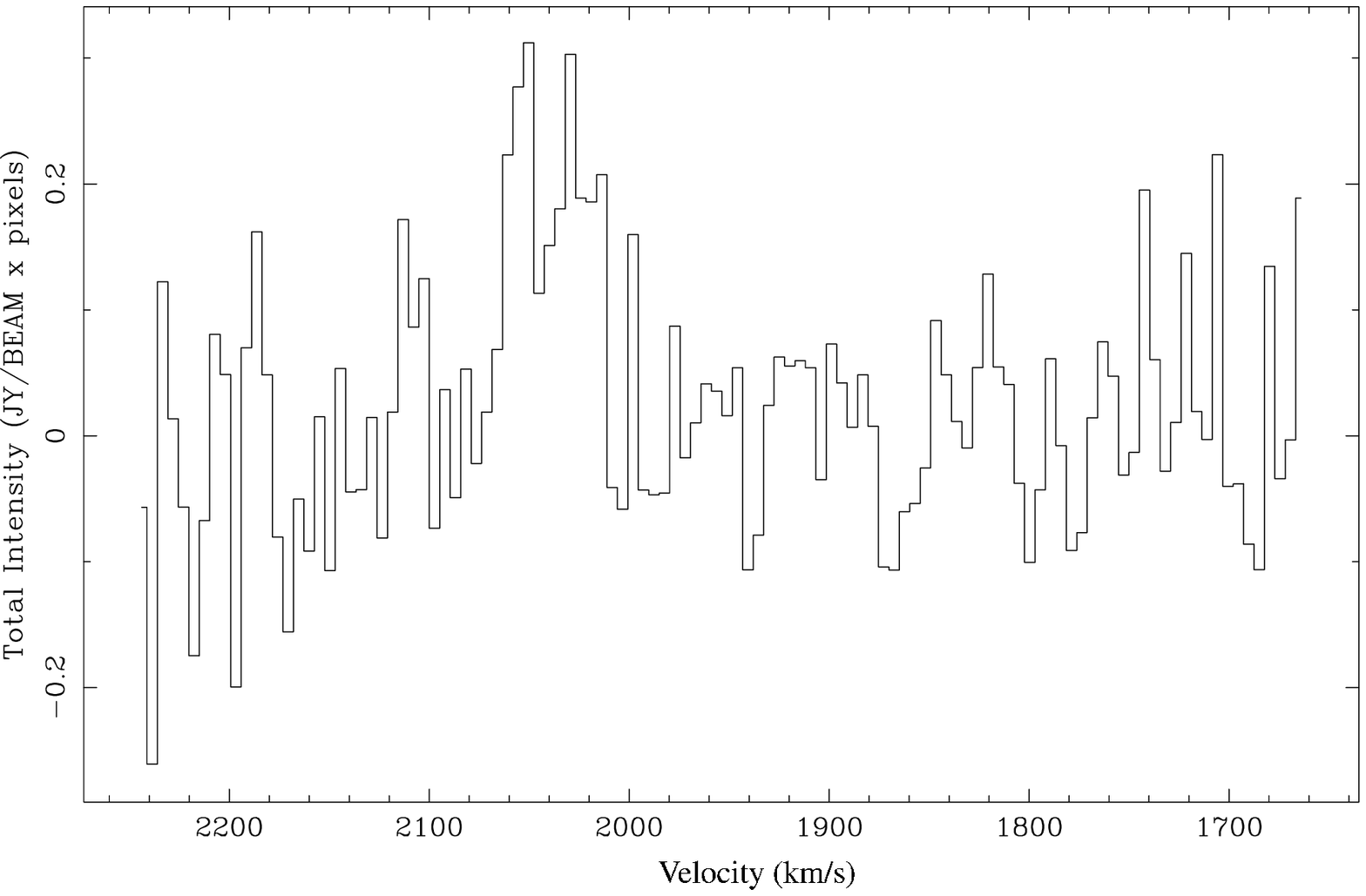}
\caption{Same as Figure~\ref{fig:gh144d1spec}, but for PWWF J1439+5847.
\label{fig:gh144d2spec}}
\end{figure}

\begin{figure}
\epsscale{0.7}
\plotone{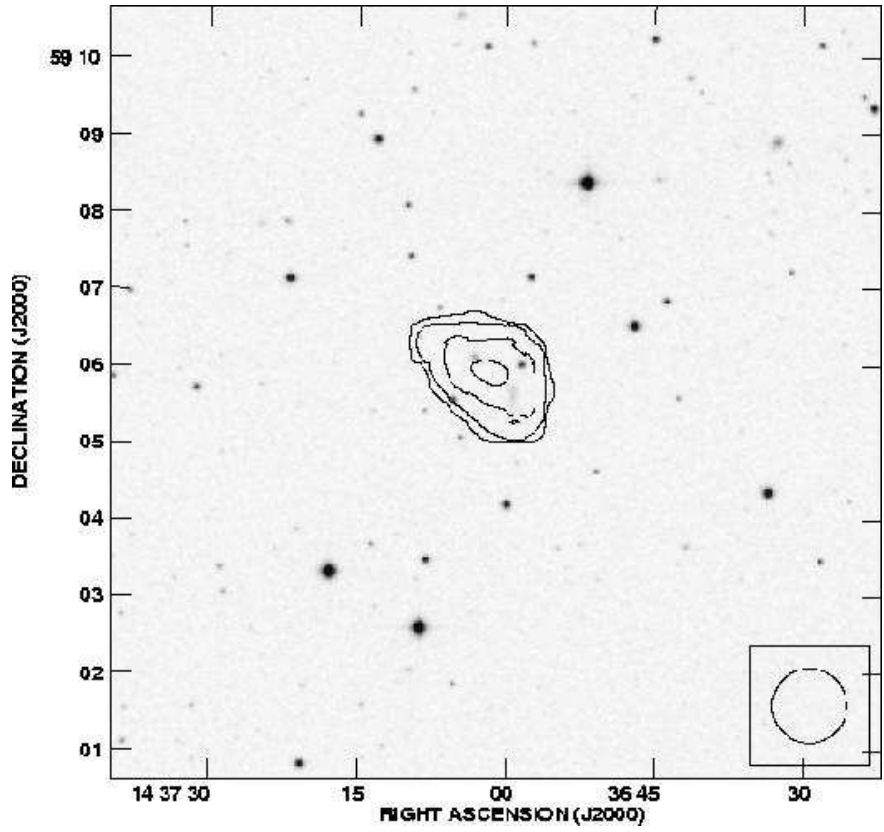}
\caption{A contour map of the total HI column density overlaid on the blue DSS-2
image of PWWF J1437+5905.  The contours are at 
0.5,1,2,3$\times$\dex{19}\cmsq.  
\label{fig:gh144d1opt}}
\end{figure}

\begin{figure}
\epsscale{0.7}
\plotone{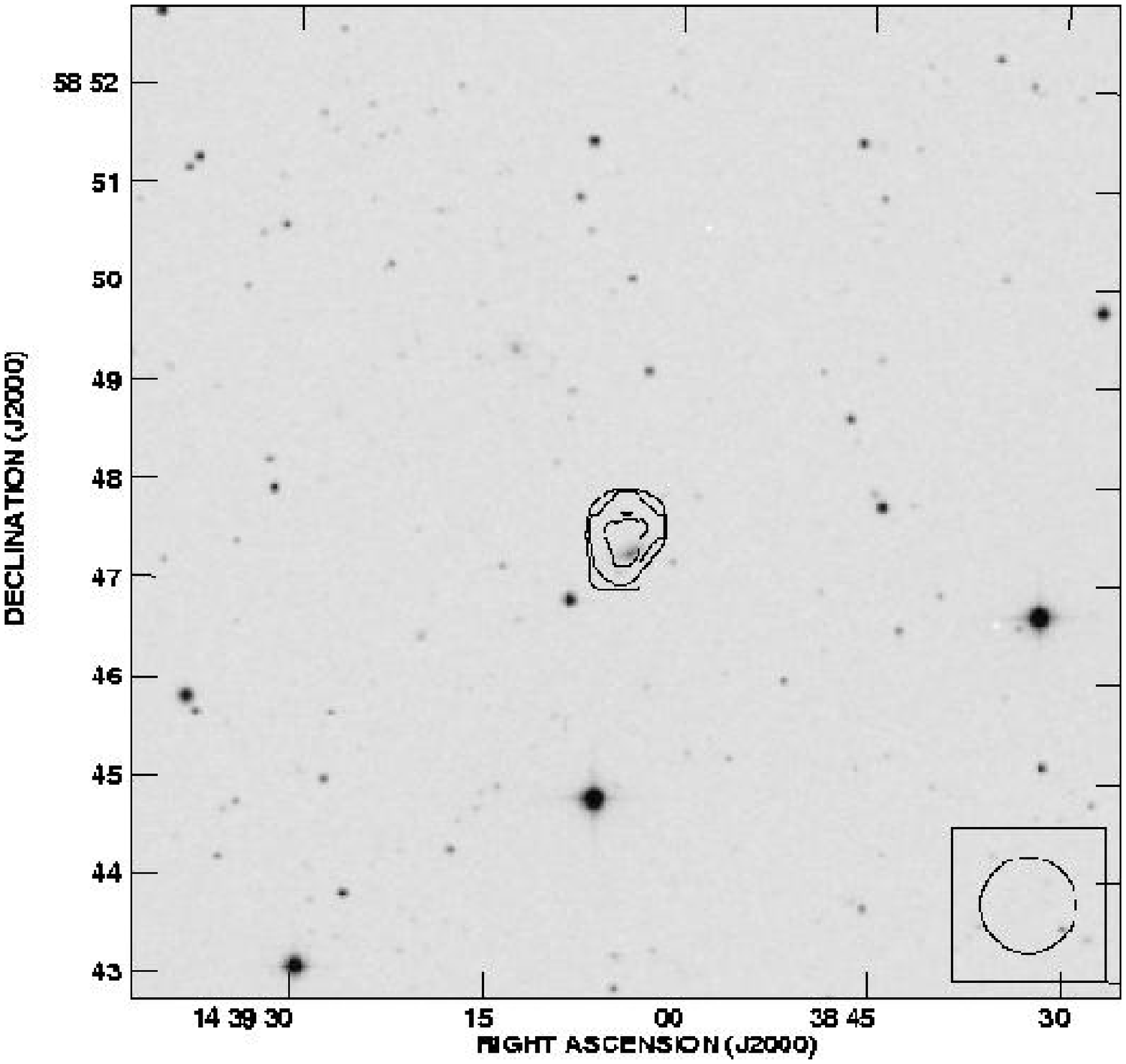}
\caption{Same as Figure~\ref{fig:gh144d1opt}, but for PWWF J1439+5847.
\label{fig:gh144d2opt}}
\end{figure}

\begin{figure}
\epsscale{0.3}
\plotone{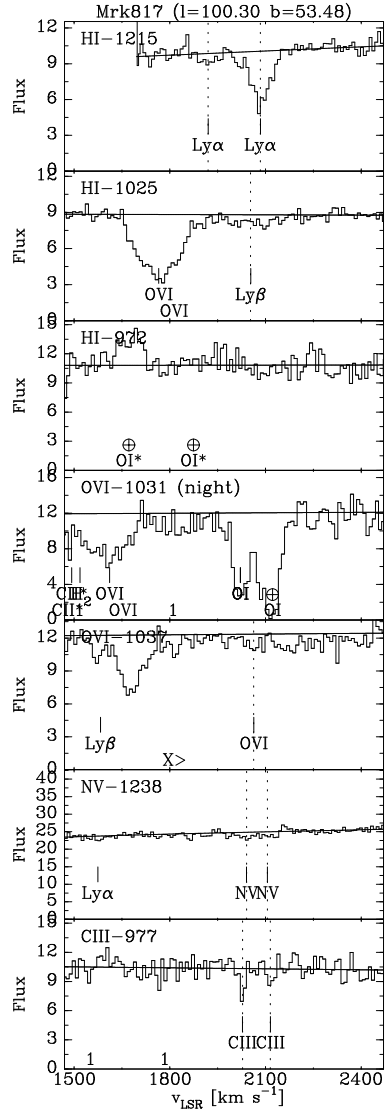}
\caption{This figure shows the Ly$\alpha$, Ly$\beta$, Ly$\gamma$, 
\OVI\l1031.926,
\OVI\l1037.617 and \CIII\l977.200 lines, for a velocity range of $\pm$500
\kms\ relative to average heliocentric velocity of GH 144 1945 \kms.  The 
flux units are as in Fig~\ref{fig:m290uv}, as are the labels and horizontal line.  
The dotted lines indicate the velocity of absorption lines associated with GH 144.
\label{fig:m817uv}}
\end{figure}

\end{document}